\documentclass[times,twocolumn]{aastex62}

\usepackage{CJK}
\usepackage{amsmath}
\usepackage{multirow}
\usepackage{cases}
\usepackage{graphicx}
\usepackage{subfigure}
\usepackage{natbib}
\usepackage{color}
\usepackage{tabularx}
\usepackage{bm}
\usepackage{threeparttable}
\usepackage{gensymb}
\usepackage{booktabs}
\usepackage{array}
\usepackage{placeins}

\newcommand{\thetae}{\theta_{\rm E}}
\newcommand{\thetas}{\theta_{*}}

\newcommand{\pie}{\pi_{\rm E}}
\newcommand{\pirel}{\pi_{\rm rel}}
\newcommand{\murel}{\mu_{\rm rel}}
\newcommand{\te}{t_{\rm E}}

\newcommand{\Mearth}{M_{\oplus}}
\newcommand{\muas}{{\rm \mu as}}
\newcommand{\eventa}{MOA-2021-BLG-145}
\newcommand{\eventb}{KMT-2026-BLG-0357}

\newcommand{\pien}{\pi_{\rm E,N}}
\newcommand{\piee}{\pi_{\rm E,E}}

\DeclareGraphicsExtensions{.pdf,.png,.jpg}

\shorttitle{}
\shortauthors{Qian et al.}

\begin{document}
\begin{CJK*}{UTF8}{gbsn}
\title{{\large Two Isolated Microlensing Events Near the Upper Shore of Einstein Desert}}

\correspondingauthor{Weicheng Zang}
\email{zangweicheng@westlake.edu.cn}

\author[0000-0003-4625-8595]{Qiyue Qian}
\affiliation{Department of Astronomy, Tsinghua University, Beijing 100084, China}
\affiliation{Department of Astronomy, Westlake University, Hangzhou 310030, Zhejiang Province, China}

\author[0000-0001-6000-3463]{Weicheng Zang}
\affiliation{Department of Astronomy, Westlake University, Hangzhou 310030, Zhejiang Province, China}

\author[0000-0001-9823-2907]{Yoon-Hyun Ryu} 
\affiliation{Korea Astronomy and Space Science Institute, Daejeon 34055, Republic of Korea}

\author{Takahiro Sumi}
\affiliation{Department of Earth and Space Science, Graduate School of Science, Osaka University, Toyonaka, Osaka 560-0043, Japan}

\author[0000-0001-5207-5619]{Andrzej Udalski}
\affiliation{Astronomical Observatory, University of Warsaw, Al. Ujazdowskie 4, 00-478 Warszawa, Poland}

\author[0000-0003-0626-8465]{Hongjing Yang}
\affiliation{Department of Astronomy, Westlake University, Hangzhou 310030, Zhejiang Province, China}

\author[0000-0002-1279-0666]{Jiyuan Zhang}
\affiliation{Department of Astronomy, Tsinghua University, Beijing 100084, China}

\author[0000-0001-8317-2788]{Shude Mao}
\affiliation{Department of Astronomy, Westlake University, Hangzhou 310030, Zhejiang Province, China}

\collaboration{(Leading Authors)}

\author[0000-0003-3316-4012]{Michael D. Albrow}
\affiliation{University of Canterbury, School of Physical and Chemical Sciences, Private Bag 4800, Christchurch 8020, New Zealand}

\author[0000-0001-6285-4528]{Sun-Ju Chung}
\affiliation{Korea Astronomy and Space Science Institute, Daejeon 34055, Republic of Korea}

\author{Andrew Gould} 
\affiliation{Max-Planck-Institute for Astronomy, K\"onigstuhl 17, 69117 Heidelberg, Germany}
\affiliation{Department of Astronomy, Ohio State University, 140 W. 18th Ave., Columbus, OH 43210, USA}

\author{Cheongho Han}
\affiliation{Department of Physics, Chungbuk National University, Cheongju 28644, Republic of Korea}

\author[0000-0002-9241-4117]{Kyu-Ha Hwang}
\affiliation{Korea Astronomy and Space Science Institute, Daejeon 34055, Republic of Korea}

\author[0000-0002-0314-6000]{Youn Kil Jung}
\affiliation{Korea Astronomy and Space Science Institute, Daejeon 34055, Republic of Korea}
\affiliation{National University of Science and Technology (UST), Daejeon 34113, Republic of Korea}

\author[0000-0002-4355-9838]{In-Gu Shin}
\affiliation{Department of Astronomy, Westlake University, Hangzhou 310030, Zhejiang Province, China}

\author[0000-0003-1525-5041]{Yossi Shvartzvald}
\affiliation{Department of Particle Physics and Astrophysics, Weizmann Institute of Science, Rehovot 7610001, Israel}

\author[0000-0001-9481-7123]{Jennifer C. Yee}
\affiliation{Center for Astrophysics $|$ Harvard \& Smithsonian, 60 Garden St.,Cambridge, MA 02138, USA}

\author[0000-0002-7511-2950]{Sang-Mok Cha} 
\affiliation{Korea Astronomy and Space Science Institute, Daejeon 34055, Republic of Korea}
\affiliation{School of Space Research, Kyung Hee University, Yongin, Kyeonggi 17104, Republic of Korea} 

\author[0000-0002-4292-9649]{Dong-Jin Kim}
\affiliation{Korea Astronomy and Space Science Institute, Daejeon 34055, Republic of Korea}

\author[0000-0003-0562-5643]{Seung-Lee Kim} 
\affiliation{Korea Astronomy and Space Science Institute, Daejeon 34055, Republic of Korea}

\author[0000-0003-0043-3925]{Chung-Uk Lee}
\affiliation{Korea Astronomy and Space Science Institute, Daejeon 34055, Republic of Korea}

\author[0009-0000-5737-0908]{Dong-Joo Lee} 
\affiliation{Korea Astronomy and Space Science Institute, Daejeon 34055, Republic of Korea}

\author[0000-0001-7594-8072]{Yongseok Lee} 
\affiliation{Korea Astronomy and Space Science Institute, Daejeon 34055, Republic of Korea}
\affiliation{School of Space Research, Kyung Hee University, Yongin, Kyeonggi 17104, Republic of Korea}

\author[0000-0002-6982-7722]{Byeong-Gon Park}
\affiliation{Korea Astronomy and Space Science Institute, Daejeon 34055, Republic of Korea}

\collaboration{(The KMTNet Collaboration)}

\author{Fumio Abe}
\affiliation{Institute for Space-Earth Environmental Research, Nagoya University, Nagoya 464-8601, Japan}

\author{David P. Bennett}
\affiliation{Code 667, NASA Goddard Space Flight Center, Greenbelt, MD 20771, USA}
\affiliation{Department of Astronomy, University of Maryland, College Park, MD 20742, USA}

\author{Aparna Bhattacharya}
\affiliation{Code 667, NASA Goddard Space Flight Center, Greenbelt, MD 20771, USA}
\affiliation{Department of Astronomy, University of Maryland, College Park, MD 20742, USA}

\author{Ian A. Bond}
\affiliation{Institute of Natural and Mathematical Sciences, Massey University, Auckland 0745, New Zealand}

\author{Kotaro Daimon}
\affiliation{Department of Earth and Space Science, Graduate School of Science, The University of Osaka, Toyonaka, Osaka 560-0043, Japan}

\author{Chihiro Ueda}
\affiliation{Department of Earth and Space Science, Graduate School of Science, The University of Osaka, Toyonaka, Osaka 560-0043, Japan}

\author{Ryusei Hamada}
\affiliation{Department of Earth and Space Science, Graduate School of Science, Osaka University, Toyonaka, Osaka 560-0043, Japan}

\author{Yuki Hirao}
\affiliation{Department of Earth and Space Science, Graduate School of Science, Osaka University, Toyonaka, Osaka 560-0043, Japan}

\author{Stela Ishitani Silva}
\affiliation{Department of Physics, The Catholic University of America, Washington, DC 20064, USA}
\affiliation{Code 667, NASA Goddard Space Flight Center, Greenbelt, MD 20771, USA}

\author{Shuma Makida}
\affiliation{Department of Earth and Space Science, Graduate School of Science, The University of Osaka, Toyonaka, Osaka 560-0043, Japan}

\author{Shota Miyazaki}
\affiliation{Department of Earth and Space Science, Graduate School of Science, Osaka University, Toyonaka, Osaka 560-0043, Japan}

\author{Yasushi Muraki}
\affiliation{Institute for Space-Earth Environmental Research, Nagoya University, Nagoya 464-8601, Japan}

\author{Seiya Nakayama}
\affiliation{Department of Earth and Space Science, Graduate School of Science, The University of Osaka, Toyonaka, Osaka 560-0043, Japan}

\author{Kansuke Nunota}
\affiliation{Department of Earth and Space Science, Graduate School of Science, Osaka University, Toyonaka, Osaka 560-0043, Japan}

\author{Ryo Ogawa}
\affiliation{Department of Earth and Space Science, Graduate School of Science, The University of Osaka, Toyonaka, Osaka 560-0043, Japan}

\author{Ryunosuke Oishi}
\affiliation{Department of Earth and Space Science, Graduate School of Science, The University of Osaka, Toyonaka, Osaka 560-0043, Japan}

\author{Hideaki Ose}
\affiliation{Department of Earth and Space Science, Graduate School of Science, The University of Osaka, Toyonaka, Osaka 560-0043, Japan}

\author{Greg Olmschenk}
\affiliation{Code 667, NASA Goddard Space Flight Center, Greenbelt, MD 20771, USA}

\author{Cl\'ement Ranc}
\affiliation{Sorbonne Universit\'e, CNRS, Institut d'Astrophysique de Paris, IAP, F-75014, Paris, France}

\author{Nicholas J. Rattenbury}
\affiliation{Department of Physics, University of Auckland, Private Bag 92019, Auckland, New Zealand}

\author[0000-0002-1228-4122]{Yuki K. Satoh}
\affiliation{College of Science and Engineering, Kanto Gakuin University, Yokohama, Kanagawa 236-8501, Japan}

\author{Daisuke Suzuki}
\affiliation{Department of Earth and Space Science, Graduate School of Science, Osaka University, Toyonaka, Osaka 560-0043, Japan}

\author{Takuto Tamaoki}
\affiliation{Department of Earth and Space Science, Graduate School of Science, Osaka University, Toyonaka, Osaka 560-0043, Japan}

\author{Sean Terry}
\affiliation{Code 667, NASA Goddard Space Flight Center, Greenbelt, MD 20771, USA}
\affiliation{Department of Astronomy, University of Maryland, College Park, MD 20742, USA}
\author{Paul J. Tristram}
\affiliation{University of Canterbury Mt.\ John Observatory, P.O. Box 56, Lake Tekapo 8770, New Zealand}

\author{Aikaterini Vandorou}
\affiliation{Code 667, NASA Goddard Space Flight Center, Greenbelt, MD 20771, USA}
\affiliation{Department of Astronomy, University of Maryland, College Park, MD 20742, USA}
\author{Hibiki Yama}
\affiliation{Department of Earth and Space Science, Graduate School of Science, Osaka University, Toyonaka, Osaka 560-0043, Japan}

\collaboration{(The MOA Collaboration)}

\author[0000-0001-7016-1692]{Przemek Mr\'{o}z}
\affiliation{Astronomical Observatory, University of Warsaw, Al. Ujazdowskie 4, 00-478 Warszawa, Poland}

\author[0000-0002-0548-8995]{Micha{\l}~K. Szyma\'{n}ski}
\affiliation{Astronomical Observatory, University of Warsaw, Al. Ujazdowskie 4, 00-478 Warszawa, Poland}

\author[0000-0002-2335-1730]{Jan Skowron}
\affiliation{Astronomical Observatory, University of Warsaw, Al. Ujazdowskie 4, 00-478 Warszawa, Poland}

\author[0000-0002-9245-6368]{Radoslaw Poleski}
\affiliation{Astronomical Observatory, University of Warsaw, Al. Ujazdowskie 4, 00-478 Warszawa, Poland}

\author[0000-0002-7777-0842]{Igor Soszy\'{n}ski}
\affiliation{Astronomical Observatory, University of Warsaw, Al. Ujazdowskie 4, 00-478 Warszawa, Poland}

\author[0000-0002-2339-5899]{Pawe{\l} Pietrukowicz}
\affiliation{Astronomical Observatory, University of Warsaw, Al. Ujazdowskie 4, 00-478 Warszawa, Poland}

\author[0000-0003-4084-880X]{Szymon Koz{\l}owski}
\affiliation{Astronomical Observatory, University of Warsaw, Al. Ujazdowskie 4, 00-478 Warszawa, Poland}

\author[0000-0002-9326-9329]{Krzysztof A. Rybicki}
\affiliation{Astronomical Observatory, University of Warsaw, Al. Ujazdowskie 4, 00-478 Warszawa, Poland}
\affiliation{Department of Particle Physics and Astrophysics, Weizmann Institute of Science, Rehovot 76100, Israel}

\author[0000-0002-6212-7221]{Patryk Iwanek}
\affiliation{Astronomical Observatory, University of Warsaw, Al. Ujazdowskie 4, 00-478 Warszawa, Poland}

\author[0000-0001-6364-408X]{Krzysztof Ulaczyk}
\affiliation{Department of Physics, University of Warwick, Gibbet Hill Road, Coventry, CV4~7AL,~UK}

\author[0000-0002-3051-274X]{Marcin Wrona}
\affiliation{Astronomical Observatory, University of Warsaw, Al. Ujazdowskie 4, 00-478 Warszawa, Poland}
\affiliation{Villanova University, Department of Astrophysics and Planetary Sciences, 800 Lancaster Ave., Villanova, PA 19085, USA}

\author[0000-0002-1650-1518]{Mariusz Gromadzki}
\affiliation{Astronomical Observatory, University of Warsaw, Al. Ujazdowskie 4, 00-478 Warszawa, Poland}

\author{Mateusz J. Mr\'{o}z}
\affiliation{Astronomical Observatory, University of Warsaw, Al. Ujazdowskie 4, 00-478 Warszawa, Poland}

\collaboration{(The OGLE Collaboration)}

\author{Yunyi Tang}
\affiliation{Department of Astronomy, Tsinghua University, Beijing 100084, China}

\author[0000-0002-1287-6064]{Zhixing Li}
\affiliation{Department of Astronomy, Westlake University, Hangzhou 310030, Zhejiang Province, China}

\author[0000-0001-5651-9440]{Yuchen Tang}
\affiliation{Department of Astronomy, Westlake University, Hangzhou 310030, Zhejiang Province, China}

\author{Xinyao Liu}
\affiliation{Department of Astronomy, Westlake University, Hangzhou 310030, Zhejiang Province, China}

\author{Dan Maoz}
\affiliation{School of Physics and Astronomy, Tel-Aviv University, Tel-Aviv 6997801, Israel}

\author[0000-0003-4027-4711]{Wei Zhu}
\affiliation{Department of Astronomy, Tsinghua University, Beijing 100084, China}

\collaboration{(The MAP Follow-up Team)}

\begin{abstract}

The Einstein desert is a deficit in the angular Einstein-radius ($\thetae$) distribution of isolated objects over the range $9~\muas <\thetae<26~\muas$. 
Although this interval contained no events when the desert was first identified, several subsequent detections have fallen within it, raising questions about its detailed structure and requiring a larger detection sample. In this paper, we report the observations and analyses of two microlensing events possibly caused by isolated giant planets or brown dwarfs whose $\thetae$ lie near the upper shore of the desert. \eventa\ has $\thetae = 23.1\pm2.0~\muas$, placing it just inside the upper boundary, and \eventb\ has $\thetae = 36.5\pm 3.7~\muas$, placing it above the desert. We also compare the $\thetae$ distributions of isolated objects and bound planetary events. The bound population shows no corresponding deficit over the same range, suggesting that the Einstein desert may be associated with the formation or subsequent evolution of the isolated population. Furthermore, some FFPs may lie above the upper shore of the Einstein desert, indicating that a more complete understanding of the FFP population requires direct lens-mass measurements.

\end{abstract}

\section{Introduction}\label{intro}
Isolated substellar objects, ranging from brown dwarfs (BDs) to free-floating planets (FFPs), are difficult to detect due to their low luminosity, leaving their origins and population properties poorly constrained. Two main formation channels, star-like and planet-like, may contribute to both BD and FFP populations. 
In star-like formation, substellar objects form through gravitational fragmentation and collapse \citep{Padoan2004,Bate2012,Luhman2012ARAA}, although how (and whether) this process extends into the planetary-mass regime remains uncertain (e.g. \citealt{Whitworth2006, Stamatellos2008, Whitworth2024}). Alternatively, objects formed in circumstellar disks may leave the system through dynamical interactions (e.g. \citealt{Rice2003,Stamatellos2009,1996Sci...274..954R, 1996Natur.384..619W, Ma2016, Gautham2025,Kaib2013}). 
These channels are not uniquely associated with a particular final mass and may overlap substantially across the planetary-to-brown-dwarf regime. Determining how their relative contributions vary with mass therefore requires an observational census of isolated substellar objects across this transition.

Gravitational microlensing \citep{Einstein1936} provides a unique means of constructing such a census because its ability to detect objects does not depend on the luminosity of the lens \citep{Mroz2017a,Gould2022_FFP_EinsteinDesert,Sumi2023}. 
A microlensing event occurs when a foreground object (the lens) crosses and deflects the light of a background star (the source) by its gravity, producing a transient magnification. 
The characteristic timescale of an event is the Einstein timescale
\begin{equation}
\te = \frac{\thetae}{\murel},
\end{equation}
where $\murel$ is the lens-source relative proper motion and $\thetae$ is the angular Einstein radius,
\begin{equation}
\label{eq:thetaE_M}
\thetae = \left(\kappa M_{\rm L}\pirel\right)^{1/2},
\end{equation}
where $M_{\rm L}$ is the lens mass, $\pirel={\rm AU}(D_{\rm L}^{-1}-D_{\rm S}^{-1})$ is the lens-source relative parallax, $D_{\rm L}$ and $D_{\rm S}$ are the distances from the Earth to the lens and source, respectively, and $\kappa=4G/(c^2{\rm au})$. Because $\te$ depends on both $\thetae$ and $\murel$, a short event timescale alone does not necessarily imply a low lens mass. By contrast, $\thetae$ is independent of the relative proper motion and is therefore more directly related to the lens mass. Although it also depends on $\pirel$, the statistics still reflects the general distribution of mass and distance.
For low-mass lenses, the small Einstein radius increases the normalized source size ($\rho=\thetas/\thetae$) making finite-source effects \citep{1994ApJ...421L..75G, Shude1994,Nemiroff1994} more likely to be measurable when the lens passes close to or across the source. In such finite-source point-lens (FSPL) events, $\rho$ can be determined from the light curve, while the angular source radius $\thetas$ can be estimated from the source color and magnitude \citep{Yoo2004}. The angular Einstein radius can therefore be measured as
\begin{equation}
\thetae = \frac{\thetas}{\rho}.
\end{equation}
FSPL events thus provide a particularly useful probe of the angular Einstein-radius distribution of isolated low-mass lenses.

Microlensing surveys have accumulated a growing sample of FSPL events, enabling the investigation of the angular Einstein-radius distribution of isolated low-mass lenses. 
\citet{KB172820} noted that all six FFP candidates known at the time were below an apparent deficit in the $\thetae$ distribution and named the sparsely populated region as ``Einstein desert". Subsequently, \citet{Gould2022_FFP_EinsteinDesert} characterized the desert more precisely as $9~\mu{\rm as} <\thetae<26~\mu{\rm as}$ using a homogeneous sample of 30 giant-source FSPL events from the 2016--2019 KMTNet data, with no events from their sample falling within this interval. The desert was interpreted as marking a transition between a low-$\thetae$ population dominated by FFPs and a higher-$\thetae$ population dominated by brown dwarfs and stars. 

Subsequent discoveries near and within the nominal desert suggest that the Einstein desert may not be an absolute gap. KMT-2022-BLG-2397, with $\thetae=24.8\pm3.6~\muas$, lies near the upper shore of the desert and is likely to be a bulge brown dwarf \citep{Gould2023KB222397}. KMT-2024-BLG-0792, interpreted as a Saturn-mass free-floating planet, has an angular Einstein radius of $\thetae=18.6 \pm 0.9~ \muas$, placing it well within the nominal desert \citep{Dong2026Sci}. 
More recently, another FFP candidate, KMT-2025-BLG-2093 with $\thetae=13.1\pm 2.8~\muas$, is located near the lower boundary of the desert \citep{Ryu2026KB252093}. These events suggest that the Einstein desert could be better viewed as a deficit in the underlying $\thetae$ distribution than as an empty interval, while its detailed structure remains uncertain. In particular, the locations and sharpness of its boundaries, and whether the transition to the higher-$\thetae$ brown-dwarf population is abrupt or gradually populated, remain open questions that require larger and more complete detections of FSPL events.

As the ongoing surveys have continued, the FSPL sample has increased, while the extension of methods including the systematic search and the follow-up observations has enabled more accurate depiction of events. Here we present the analysis of two FSPL events near the upper shore of the Einstein desert, \eventa\ on the inner side with $\thetae=23.1\pm 2.0~\muas$ and \eventb\ on the outer side with $\thetae=36.5\pm 3.7~\muas$. The details of the observations are shown in Section \ref{obser}, and the modeling processes are described in Section \ref{model}. We analyze the source and lens properties in Section \ref{lens} and discuss the $\thetae$ distribution of isolated objects and bound planetary events in Section \ref{dis}.

\section{Observations}\label{obser}

\begin{figure}[ht!]
    \centering
    \includegraphics[scale=0.35]{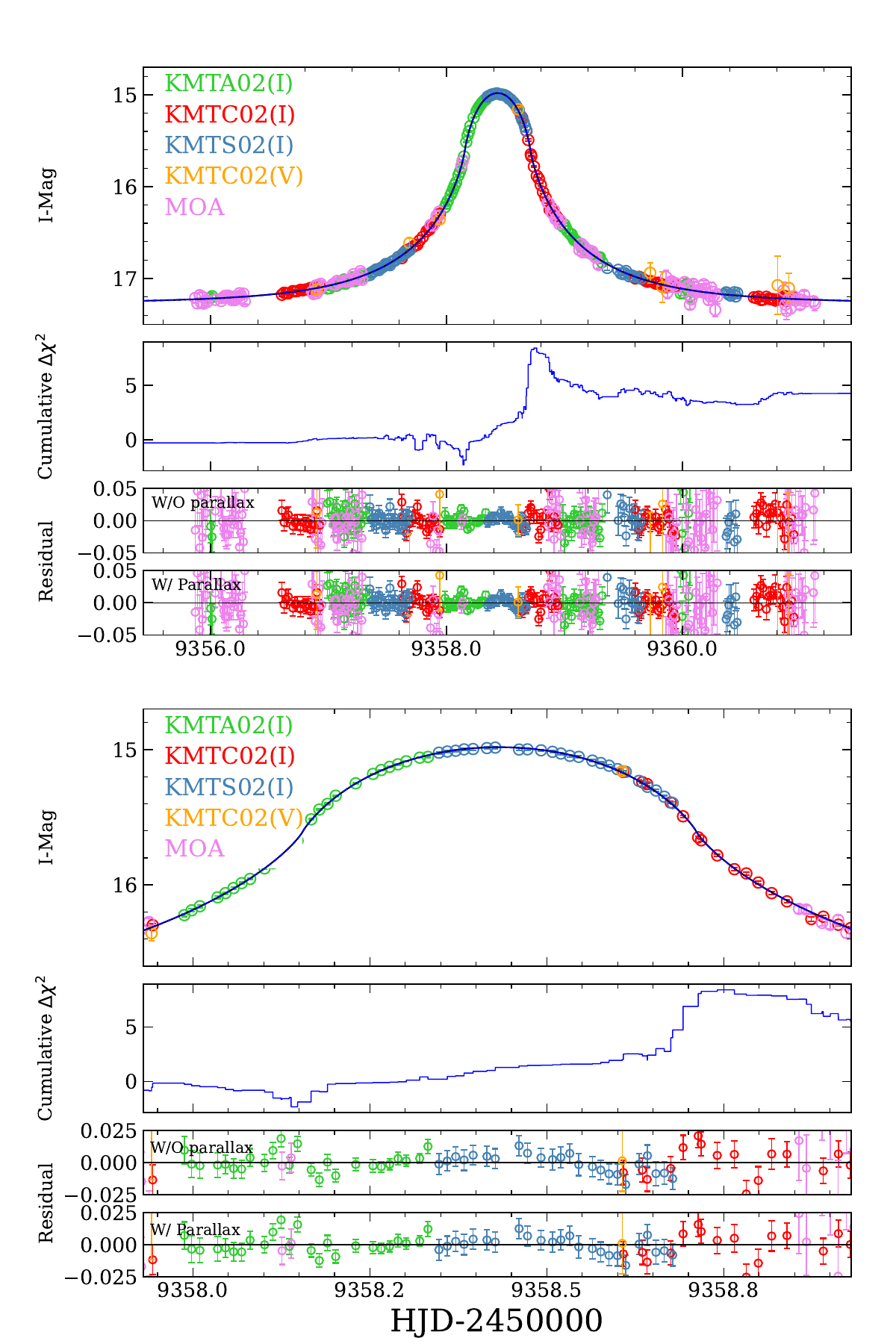}
    \caption{The observed data points, the best-fit models including non-parallax FSPL and with-parallax FSPL of the \eventa\ light curve, the cumulative $\Delta\chi^2$ of two models, and the residuals of data to the models. The magnitude system are scaled to the KMTC02 instrumental magnitude in I-band.}
    \label{fig:KB210950lc}
\end{figure}

Figure~\ref{fig:KB210950lc} displays the observed data for \eventa, which is located at equatorial coordinates $(\alpha, \delta)_{\text{J2000}}$ = (17:56:19.30, $-$29:07:05.02), corresponding to Galactic coordinates $(\ell, b) = (1.0403, -2.1032)$. It was first discovered in real time by the second phase of Microlensing Observations in Astrophysics (MOA-II, \citealt{Sako2008}) at 18:59 on May 23, 2021 ($\rm HJD^\prime = 9358.29$, HJD$^\prime = \mathrm{HJD} -2450000$), and was designated as \eventa. About 7.5 hours later, it was independently identified by the Korea Microlensing Telescope Network (KMTNet, \citealt{KMT2016}) \texttt{AlertFinder} system \citep{KMTAF} and labeled as KMT-2021-BLG-0950. This event was also recovered by the systematic search for FFPs in the KMTNet full-frame images \citep{KMTFFP1}, using all images collected in 2021.

\begin{figure}[ht!]
    \centering
    \includegraphics[scale=0.35]{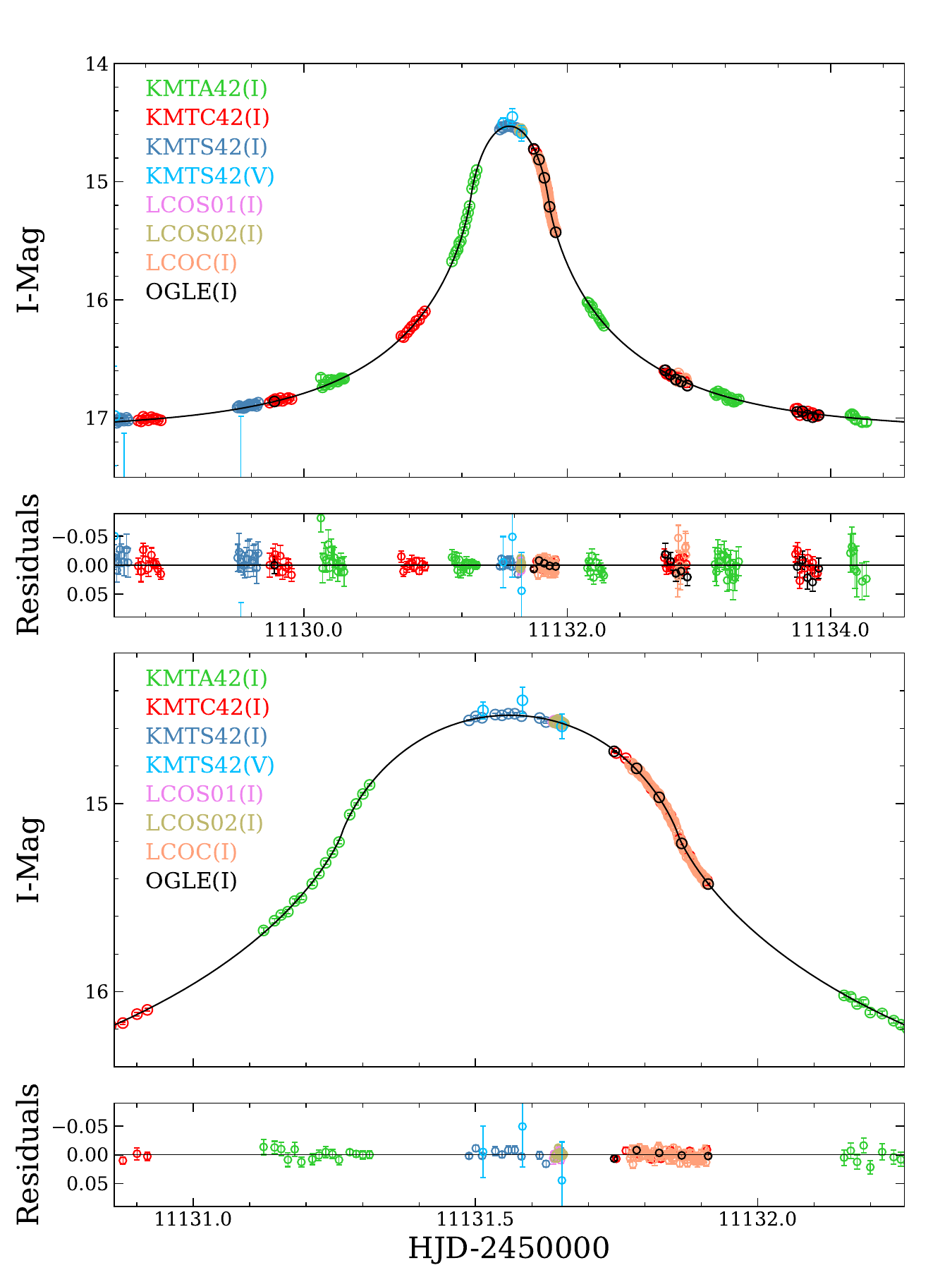}
    \caption{The observed data points, the best-fit model (FSPL) of the \eventb\ light curve, and the residual of data to the model. The lower panel provides a zoom-in view of the event. Different colors represent data from different observational sites and bands that have their individual $(f_{\rm s},~f_{\rm b})$. The magnitude system are scaled to the OGLE-IV in $I$-band magnitude.}
    \label{fig:KB260357lc}
\end{figure}

Figure~\ref{fig:KB260357lc} shows the observed data for \eventb. The event is located at equatorial coordinates $(\alpha, \delta)_{\text{J2000}}$ = (17:54:23.48, $-$28:14:00.24), corresponding to Galactic coordinates $(\ell, b) = (1.5915, -1.2919)$. It was first announced by the KMTNet \texttt{AlertFinder} system at 01:43 on March 31, 2026 ($\rm HJD^\prime = 11130.57$), about one day before the peak magnification. The event was later independently identified by the Early Warning System \citep{Udalski1994,Udalski2003} of the Optical Gravitational Lensing Experiment \citep[OGLE,][]{OGLEIV} as OGLE-2026-BLG-0156. At $\rm HJD^\prime = 11131.36$, it was also flagged by the KMTNet \texttt{HighMagFinder} system as a candidate high-magnification event \citep{KB210171}. Following the alert, the event was monitored around the peak by the Las Cumbres Observatory Global Telescope (LCOGT; \citealt{LCOGT}) under the LCOGT Key Project (Program ID: KEY2023B-003) for KMTNet high-magnification events \citep{KB200414}.

KMTNet obtained observations with its three 1.6~m telescopes, each equipped with a $4\,{\rm deg}^2$ field-of-view (FoV) camera, located at observatories in Chile (KMTC), South Africa (KMTS), and Australia (KMTA) \citep{KMT2016}. The MOA survey utilized a 1.8~m telescope in New Zealand equipped with a $2.2\,{\rm deg}^2$ FoV camera \citep{Sako2008}, while OGLE conducted observations with its 1.3~m telescope in Chile using a $1.4\,{\rm deg}^2$ FoV camera \citep{OGLEIV}. The LCOGT data were obtained with two 1.0~m telescopes in South Africa (LCOS01 and LCOS02) and one 1.0~m telescope in Chile (LCOC). For \eventa, the observing cadences were $\Gamma = 3~\mathrm{hr}^{-1}$ for KMTA and KMTS, $\Gamma = 2~\mathrm{hr}^{-1}$ for KMTC, and $\Gamma \sim 4~\mathrm{hr}^{-1}$ for the MOA survey. For \eventb, the observing cadences were $\Gamma = 3.6~\mathrm{hr}^{-1}$ for KMTA and KMTS, $\Gamma = 2.4~\mathrm{hr}^{-1}$ for KMTC, while the OGLE cadence was about $\Gamma = 1~\mathrm{hr}^{-1}$.

Most KMTNet and OGLE images were obtained in the $I$ band, whereas MOA primarily used its broad MOA-Red filter, which approximately corresponds to the combination of the Cousins $R$ and $I$ bands. KMTNet and MOA also obtained a fraction of $V$-band images to measure the source star colors. The LCOGT images were obtained solely in the $I$ band. In addition to the primary-band observations, we include the KMTC $V$-band data for \eventa\ and the KMTS $V$-band data for \eventb\ in the light-curve modeling, as both datasets have sufficient signal-to-noise ratios (SNRs) near the event peaks. 

The photometry of both \eventa\ and \eventb\ was extracted using the Difference Image Analysis (DIA) technique \citep{Tomaney1996,Alard1998}. Specifically, the KMTNet and LCOGT data were reduced with the \texttt{pySIS} pipeline \citep{pysis,Yang_TLC,Yang_TLC2}, the OGLE data with the \citet{Wozniak2000} pipeline, and the MOA data with the \citet{Bond2001} pipeline. The photometric uncertainties from each DIA pipeline were re-normalized following the procedure described by \citet{MB11293}, such that the reduced chi-squared ($\chi^2/{\rm dof}$) for each dataset is unity, where ${\rm dof}$ represents the degree of freedom.

\section{Light-curve Analysis}\label{model}

For both events, the light curves exhibit pronounced finite-source effects around the peak, resulting in broadened and flattened peaks relative to the point-source model. We therefore model both events using the FSPL model, which is characterized by four parameters: $t_0$, $\te$, $u_0$, and $\rho$. The first three are the Paczy\'{n}ski parameters \citep{Paczynski1986}: $t_0$ is the time of the closest approach between the source and the lens, $u_0$ is the impact parameter normalized to the angular Einstein radius, and $\te$ is the Einstein timescale.The fourth parameter is the normalized source radius, defined as the ratio of the angular source radius, $\thetas$, to the angular Einstein radius, $\rho=\thetas/\thetae$. 

For each data set, we introduce two flux parameters, $(f_{\rm S}, f_{\rm B})$, representing the source flux and blended flux, respectively. The FSPL magnification is computed using the \texttt{VBBinaryLensing} package \citep{Bozza2010,VBMicrolensing2025}.  The $u_0$ of two events are both close to 0, so we do not impose any restrictions on its sign in the modeling process.

In addition, the modeling includes the limb-darkening effect, which causes the finite-source magnification to differ among passbands. In general, the source color can be determined by the regression of the $I$- and $V$-band light curves outside the finite-source region, where the magnification is insensitive to limb darkening. The source color is then used to infer the source properties and determine the corresponding limb-darkening coefficients. However, the two short-timescale events presented here lack $V$-band data with sufficient signal-to-noise ratios outside the finite-source region to accurately constrain the source color. Instead, the available $V$-band data are obtained primarily during the finite-source phase, making the modeling dependent on the adopted limb-darkening coefficients. 

We therefore iteratively perform the light-curve modeling and source-property analysis (Section~\ref{lens}), adopting as the initial guess the typical source color for stars of similar brightness. We adopt the linear limb-darkening law \citep{An2002,Claret2011}, assuming solar metallicity, a microturbulent velocity of $1~\rm km\,s^{-1}$, and $\log g = 2.5$, following \citet{Claret2011}. The final adopted coefficients are $u_I=0.63$, $u_V=0.83$, and $u_{\rm MOA}=(u_I+u_R)/2=0.68$ for \eventa, and $u_I=0.58$ and $u_V=0.78$ for \eventb. 

Table~\ref{table:bestparams} lists the resulting FSPL parameters for both events, derived from a Markov chain Monte Carlo (MCMC) exploration of the $\chi^2$ surface using the \texttt{emcee} sampler \citep{emcee2,emcee}. The reported values correspond to the posterior medians, with the uncertainties representing the 68\% credible intervals. Owing to the dense coverage of the event peaks, the normalized source radius is tightly constrained for both events, with a precision at $\sim 1\%$: $\rho = 0.2397 \pm 0.0025$ for \eventa\ and $\rho = 0.1783 \pm 0.0014$ for \eventb.

\begin{figure}[ht!]
    \centering
    \includegraphics[scale=0.6]{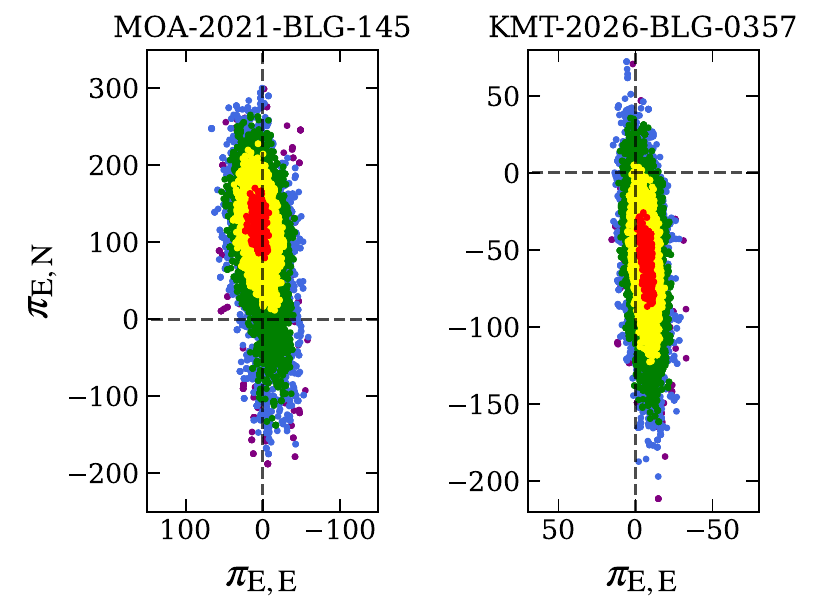}
    \caption{The 2D distribution of $\piee$ and $\pien$ of \eventa\ (left panel) and \eventb\ (right panel). The color of points represent the $\Delta \chi^2=\chi^2_{\rm chain}-\chi^2_{\rm best}<n^2$, where the (red, yellow, green, blue, purple) is corresponding to $n=$(1, 2, 3, 4, 5). The black dash lines are $\piee=0$ and $\pien=0$.}
    \label{fig:parallax}
\end{figure}

In addition, we investigate whether the microlensing parallax effect \citep{Gould1992,Gould2000} can be constrained from the light curves. Specifically, we consider both the annual parallax effect caused by Earth's orbital motion and the terrestrial parallax effect \citep{HW95, Holz1996} caused by observations from different sites on Earth. For \eventa, the MOA and KMTA data, as well as the KMTC and KMTS data, provide simultaneous observations during the finite-source effects, potentially enabling a measurement of the terrestrial parallax effect. The microlensing parallax vector is defined as
\begin{equation}
    \boldsymbol{\pi}_{\rm E}
    =
    \frac{\pi_{\rm rel}}{\theta_{\rm E}}
    \frac{\boldsymbol{\mu}_{\rm rel}}{\mu_{\rm rel}},
\end{equation}
such that its magnitude is $\pi_{\rm E}=\pi_{\rm rel}/\theta_{\rm E}$ and its direction is parallel to the lens-source relative proper motion. We fit the north and east components of the parallax vector in equatorial coordinates, $\pien$ and $\piee$. 

For \eventa, including the parallax effect improves the fit by $\Delta\chi^2 = 4.2$. The left panel of Figure~\ref{fig:parallax} shows the 2D distribution of the two parallax components, with mean values of $\pien \sim 100$ and $\piee \sim 0$. We further examine the distribution of $\Delta\chi^2$, as shown in Figure~\ref{fig:KB210950lc}. Most of the improvement arises from the interval covered by the simultaneous KMTC and KMTS observations near the peak. However, even after including the parallax effect, the correlated residuals in the KMTC and KMTS data are not fully eliminated, suggesting the presence of potential systematics. Furthermore, following the procedures described in Section~\ref{lens} to estimate the lens mass and distance, the mean parallax values would imply a very low-mass lens of $M_{\rm L}\sim 10~\Mearth$ at a very nearby distance of $D_{\rm L}\sim 0.4~\rm kpc$, which is unlikely. Therefore, we adopt only the 3-$\sigma$ upper limit on parallax, $\pie < 266$, to infer a lower limit on the lens mass.

For \eventb, the 2D distribution of $\boldsymbol{\pi}_{\rm E}$ is illustrated in the right panel of Figure~\ref{fig:parallax}. The mean value of $\pie$ is about 60, which would also suggest a very nearby lens ($D_{\rm L}\sim 0.5~\rm kpc$). However, the improvement in $\chi^2$ is only 2.7. Therefore, we also adopt the 3-$\sigma$ upper limit on the parallax, $\pie < 162$.

\begin{table}[htbp]
\centering
\renewcommand{\arraystretch}{1.2}
\caption{FSPL Parameters for the Two Events}
\begin{tabular}{l|c|c}
\hline\hline
Parameter & \eventa & \eventb \\
\hline
$t_0$ (HJD') & $9358.43302 \pm 0.00041$& $11131.56057 \pm 0.00047$\\
$t_{\mathrm{E}}$ (day) & $1.1658 \pm 0.0097$& $1.667 \pm 0.011$\\
$u_0~(10^{-2})$ & $2.2 \pm 2.0$& $0.009 \pm 0.081$\\
$\rho$ & $0.2397 \pm 0.0025$& $0.1783 \pm 0.0014$\\
$I_{\mathrm{S}}$ & $17.419 \pm 0.015$& $17.5799 \pm 0.0092$\\
$V_{\mathrm{S}}$ & $20.188 \pm 0.015$& $21.5223 \pm 0.0092 $\\
\hline
\end{tabular}
{\textit{Note.} HJD$^\prime$ = HJD $-$ 2450000. The source magnitudes, $I_{\rm S}$ and $V_{\rm S}$, are reported in the KMTC02 instrumental system for \eventa\ and the KMTS42 instrumental system for \eventb.}
\label{table:bestparams}
\end{table}

\section{Source and Lens Properties}\label{lens}

The angular Einstein radius $\thetae$ and the relative proper motion $\murel$ of the lens can be derived by 
\begin{equation}
\label{eq:thetae}
    \thetae=\frac{\thetas}{\rho},~~\murel=\frac{\thetae}{t_{\rm E}}.
\end{equation}
The fitting of the FSPL model constrains $\rho$ and $\te$, while both the limb-darkening parameters in the modeling process and $\thetas$ estimation require the intrinsic magnitude and color of the source. 

The light-curve modeling (Section~\ref{model}) yields the apparent source magnitudes and colors. Toward the Galactic bulge, however, these quantities are affected by interstellar extinction and reddening. The extinction and reddening are commonly estimated using the red clump (RC) stars, whose intrinsic color and magnitude are well calibrated \citep{Bensby2013,Nataf2013}. The procedure is to construct a color-magnitude diagram (CMD) of stars in the vicinity of the event, determine the centroid of the RC following the method of \citet{Nataf2013}, and compare the source position with the RC centroid. The intrinsic source color and magnitude are then obtained from their offsets relative to the RC. Using the $(V-I,I)$ CMD as an example, the intrinsic source color and magnitude are given by
\begin{equation}
    (V-I,I)_{\rm S,0} = (V-I,I)_{\rm S} - (V-I,I)_{\rm RC} + (V-I,I)_{\rm RC,0},
\end{equation}
where the subscripts ``S'' and ``RC'' denote the source and the RC centroid, respectively, and the subscript ``0'' denotes intrinsic (de-reddened) quantities. 

Applying the color/surface-brightness (CSB) relation of \citet{Adams2018},
\begin{equation}
    \log (2\theta_*)=\sum_{n=0}^{2} c_n(V-I)^n-0.2I,
\end{equation}
where $V-I$ and $I$ may be replaced by the corresponding color and magnitude in the photometric bands used, and the coefficients $c_n$ for different stellar types and bandpasses are taken from \citet{Adams2018}. This relation yields the angular source radius, $\theta_*$.

The intrinsic source color can also be used to estimate the effective temperature, $T_{\rm eff}$, through the color-temperature relations of \citet{Houdashelt2000}. The derived $T_{\rm eff}$ is then used to determine the limb-darkening coefficients and to iteratively refine the light-curve modeling described in Section~\ref{model}.

\subsection{\eventa}

\begin{figure}[ht!]
    \centering
    \includegraphics[scale=0.6]{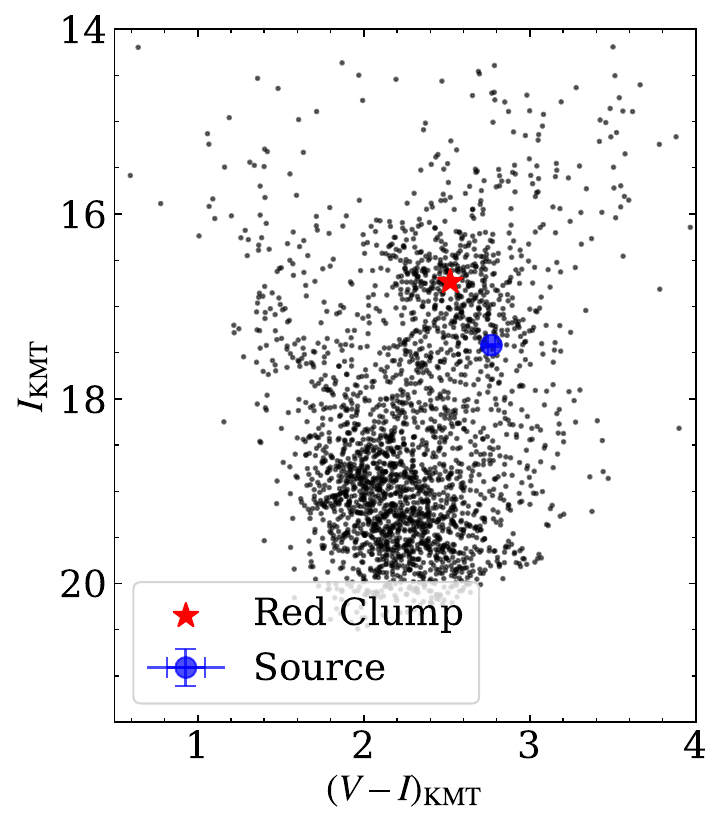}
    \caption{The instrumental CMD of \eventa. Black dots represent the KMTNet field stars within a $3^{\prime}\times3^{\prime}$ region centered on the event. The red asterisk are the estimated RC centroid, and the blue dots illustrate the source color and magnitude.}
    \label{fig:KB210950cmd}
\end{figure}

We construct the CMD from KMTNet self-constructed $I$- and $V$-band catalogs, using the KMTC datasets as same as those adopted for the source-color measurement. For each band, we extract a stellar catalog from a $3^{\prime}\times3^{\prime}$ stamp image centered on the event and cross-match the two catalogs to obtain the $(V-I,I)$ measurements of stars in the field. Following the procedure described above, we determine the RC centroid to be $(V-I,I)_{\rm RC}=(2.521\pm0.015,~16.734\pm 0.047)$. The source color and magnitude inferred from the light-curve modeling in Section~\ref{model} is $(V-I,I)_{\rm S}=(2.769\pm0.021,~17.419\pm 0.015)$, placing it redward and fainter than the RC centroid. Figure \ref{fig:KB210950cmd} displays the CMD, the RC centroid, and the source of \eventa. 

With the RC centroid color and magnitude to correct the extinction and reddening, we obtain the source intrinsic color-magnitude of $(V-I,I)_{\rm S,0}=(1.309\pm0.043,~15.070\pm 0.064)$, indicating that the source is a red giant. The intrinsic color corresponds to an effective temperature of $T_{\rm eff}\sim 4225~{\rm K}$. Applying the CSB relation then yields an angular source radius of
\begin{equation}
\theta_* = 5.53\pm0.47~\mu{\rm as}.
\end{equation}
Combining this result with $\rho=0.2397\pm0.0025$ and $t_{\rm E}=1.1658\pm0.0097~{\rm day}$ from the final iterated FSPL solution, we derive
\begin{equation}
\theta_{\rm E}=23.1 \pm 2.0~\mu{\rm as},
\qquad
\mu_{\rm rel}=7.22 \pm 0.63~{\rm mas~yr^{-1}}.
\end{equation}
The central value of $\theta_{\rm E}$ lies within the nominal Einstein-desert interval, $9~\mu{\rm as}<\theta_{\rm E}<26~\mu{\rm as}$, and is $\sim 3~\mu{\rm as}$ below its upper shore approximately at the level of $1.5\sigma$. 

According to Eq.~\ref{eq:thetaE_M}, the lens mass is 
\begin{equation}
\label{eq:thetaE_M_a}
    M_{\rm L} = \frac{\thetae^2}{\kappa\pirel} = 1.36~M_{\rm Jup} \left( \frac{\pirel}{0.05~\rm mas} \right)^{-1}.
\end{equation}
If the lens is in the bulge, for example, $D_{\rm L}=7.0~\rm kpc,\ D_{\rm S}=8.0~\rm kpc$, the mass is $M_{\rm L} = 3.8~M_{\rm Jup}$. If the lens is in the disk, for example, $D_{\rm L}=4.0~\rm kpc,\ D_{\rm S}=8.0~\rm kpc$, the mass is $M_{\rm L} = 0.54~M_{\rm Jup}$. These estimates indicate that the lens is likely a gas-giant FFP. Adopting the 3-$\sigma$ limit of parallax, the lens mass and distance are restricted to be $M\gtrsim 3.6~\Mearth$ and $D_{\rm L}\gtrsim 160$~pc.

\begin{table}[htbp]
\centering
\renewcommand{\arraystretch}{1.2}
\caption{CMD parameters, source properties and derived $\thetae$ and $\murel$}
\begin{tabular}{l|c|c}
\hline\hline
Parameter & \eventa & KMT-2026-BLG-0357 \\
\hline
$(r-z)_{\rm RC}$ & $\cdots$ & $2.663\pm0.014$ \\
$z_{\rm RC}$ & $\cdots$ & $16.845\pm 0.055$ \\
$(V-I)_{\rm RC}$ & $2.521 \pm 0.015$ & $3.952 \pm 0.034$ \\
$I_{\rm RC}$ & $16.734 \pm 0.047$ & $17.661 \pm 0.056$ \\
$(V-I)_{\rm RC,0}$ & $1.06\pm 0.03$ & $1.06\pm 0.03$ \\
$I_{\rm RC,0}$ & $14.39 \pm 0.04$ & $14.38 \pm 0.04$ \\
$(V-I)_{\rm S}$ & $2.769\pm0.021$& $3.942\pm0.013$\\
$I_{\rm S}$ & $17.418 \pm 0.015$& $17.5799\pm0.0092$\\
$(V-I)_{\rm S,0}$ & $1.308 \pm 0.040$& $1.050\pm0.047$\\
$I_{\rm S,0}$ & $15.080 \pm 0.063$& $14.301\pm0.069$\\
$\theta_{*}$ ($\muas$) & $5.53 \pm 0.47$& $6.51\pm0.66$\\
$\theta_{\rm E}$ ($\muas$) & $23.1 \pm 2.0$& $36.5\pm 3.7$\\
$\mu_{\rm rel}$ (mas yr$^{-1}$) & $7.22 \pm 0.63$& $7.99\pm0.82$\\
\hline
\end{tabular}
\label{table:cmd}
\end{table}

\subsection{\eventb}

\begin{figure*}[ht!]
    \centering
    \includegraphics[scale=0.4]{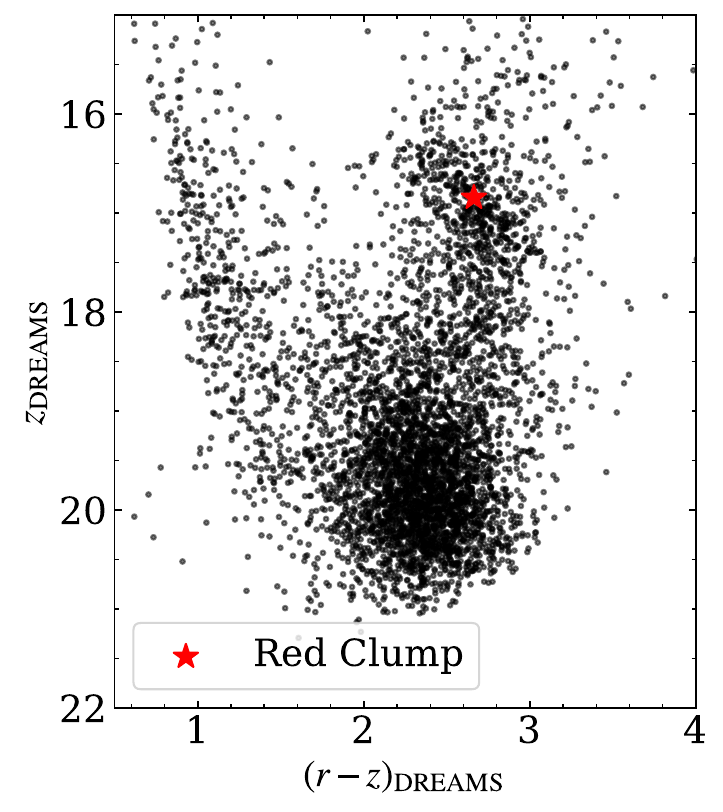}
    \includegraphics[scale=0.4]{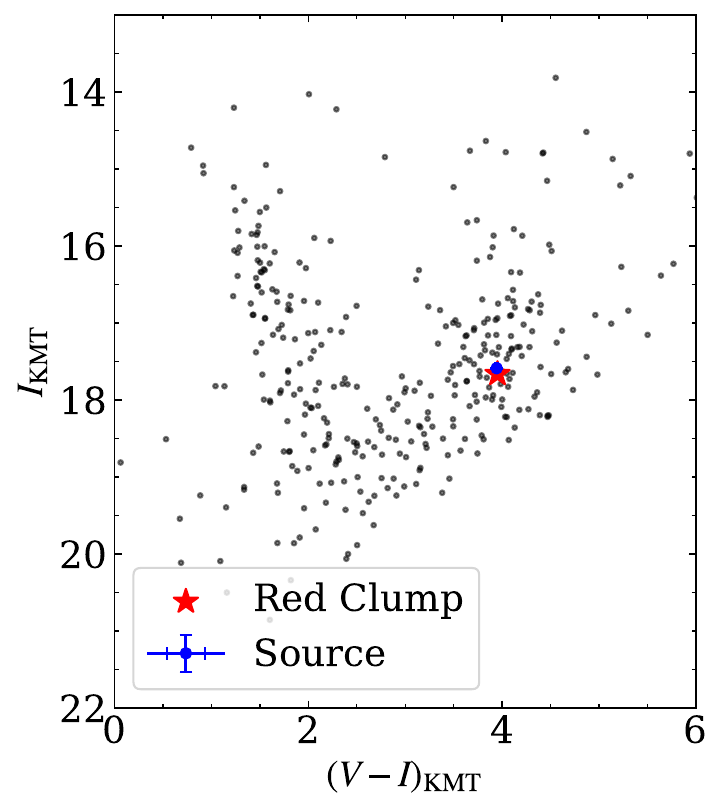}
    \includegraphics[scale=0.4]{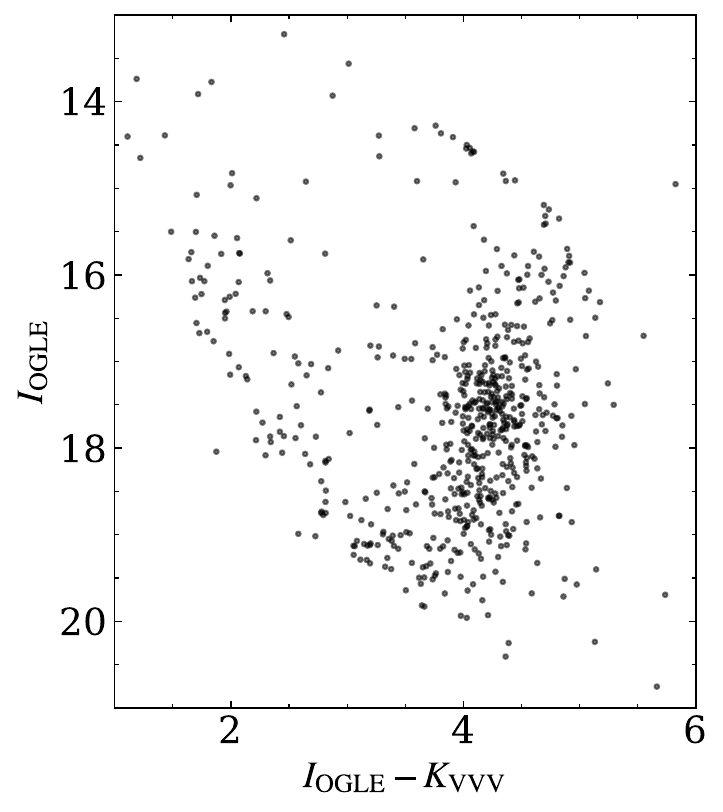}
    \caption{The CMDs of \eventb, constructing from the DREAMS (left panel) within $3^{\prime}$ radius, KMTNet (middle panel) within a $2^{\prime}\times2^{\prime}$ region and, OGLE-IV and VVV (right panel) within a $2^{\prime}\times2^{\prime}$ region. In the left panel, the red asterisk marks
    the RC centroid obtained from the modeling of RC distribution. The RC centroid is converted from $(r-z, z)$ to $(V-I, I)$ and marked by the red asterisk in the middle panel. The blue dot shows the color and magnitude of the source.} 
    \label{fig:KB260357cmd}
\end{figure*}

\eventb\ is located in a high-extinction field. Due to the detection limit, CMDs constructed from the KMTNet and VVV catalog \citep{VVV,VVV_cat} do not extend sufficiently below the RC to sample the full clump distribution, as illustrated in the middle and right panels of Figure \ref{fig:KB260357cmd}. 
This limitation is overcome using the substantially deeper catalog from the DECam Rogue Earths and Mars Survey (DREAMS; \citealt{DREAMS_DR1}). 
Starting from 2025, the DREAMS survey conducts minute-cadence observations of a $5~{\rm deg}^{2}$ region toward the Galactic bulge using DECam mounted on the 4 m Blanco Telescope at CTIO. The DREAMS DR1 published a deep stellar catalog on the $r-$ and $z-$bands. In the \eventb\ field, the catalog extends approximately four magnitudes below the RC in the $z$ band, allowing the RC to be clearly identified, as illustrated in the left panel of Figure \ref{fig:KB260357cmd}. 

We apply the RC-centroid procedure as above to the DREAMS $(r-z, z)$ CMD within a $3^{\prime}$ radius of the event, and estimate the apparent color and magnitude of the RC centroid as $(r-z, z)_{\rm RC}=(2.663\pm 0.014, 16.845\pm 0.055)$. To transform the RC centroid to the $(V-I,I)$ system used for the source measurement, we cross-match stars between the DREAMS and KMTNet catalogs and derive empirical color-color relations between $r-z$ and $V-I$, and between $r-z$ and $I-z$. Applying these relations yields $(V-I,I)_{\rm RC}=(3.952\pm0.034,~17.661\pm 0.056)$. The source color and magnitude inferred from the light-curve modeling in Section~\ref{model} is $(V-I,I)_{\rm S}=(3.942\pm0.013,~17.5799\pm 0.0092)$, lying nearly coincident with the RC centroid but slightly bluer and brighter. Using the RC centroid as a local reference to correct for extinction and reddening, we obtain the intrinsic source color and magnitude $(V-I,I)_{\rm S,0}=(1.050\pm0.047,~14.301\pm 0.069)$, suggesting that the source is a red giant.

The intrinsic color corresponds to an effective temperature of $T{\rm eff}\sim 4678~{\rm K}$ and the angular source radius is
\begin{equation}
\theta_* = 6.51\pm0.66~\mu{\rm as}.
\end{equation}
Combining this result with $\rho=0.1783\pm0.0014$ and $t_{\rm E}=1.667\pm0.011~{\rm day}$ from the final iterated FSPL solution, we obtain
\begin{equation}
\theta_{\rm E}=36.5 \pm 3.7~\mu{\rm as},
\qquad
\mu_{\rm rel}=7.99 \pm 0.82~{\rm mas~yr^{-1}}.
\end{equation}
In contrast to \eventa, whose central value of $\theta_{\rm E}$ is within the Einstein desert, the measured $\theta_{\rm E}$ is clearly above the upper shore $26~\mu{\rm as}$. 

According to Eq.~\ref{eq:thetaE_M}, the lens mass is 
\begin{equation}
\label{eq:thetaE_M_b}
    M_{\rm L} = \frac{\thetae^2}{\kappa\pirel} = 3.48~M_{\rm Jup} \left( \frac{\pirel}{0.05~\rm mas} \right)^{-1}.
\end{equation}
If the lens is in the bulge, for example, $D_{\rm L}=7.0~\rm kpc,\ D_{\rm S}=8.0~\rm kpc$, the mass is $M_{\rm L} = 9.8~M_{\rm Jup}$. If the lens is in the disk, for example, $D_{\rm L}=4.0~\rm kpc,\ D_{\rm S}=8.0~\rm kpc$, the mass is $M_{\rm L} = 1.4~M_{\rm Jup}$. Therefore, the lens is likely a gas-giant FFP or an isolated brown dwarf. Adopting the 3-$\sigma$ limit of parallax, the lens mass and distance are restricted to be $M\gtrsim 9~\Mearth$ and $D_{\rm L}\gtrsim 166$~pc.

\section{Discussion: $\thetae$ Distributions for Isolated and Bound Substellar Objects}\label{dis}

In this work, we have presented two FSPL events produced by apparently isolated low-mass lenses, and both have intermediate angular Einstein radii near the upper shore of the Einstein desert at $9~\muas <\thetae<26~\muas$. \eventa\ has $\thetae=23.1\pm2.0~\muas$ and thus lies just inside the nominal upper boundary at $26~\muas$, whereas \eventb\ has $\thetae=36.5\pm 3.7~\muas$ and lies above it. Besides the sporadic detections on the desert after its discovery, the two events further enlarge the sample near the upper shore of the Einstein desert.

\begin{figure}[ht!]
    \centering
    \includegraphics[scale=0.4]{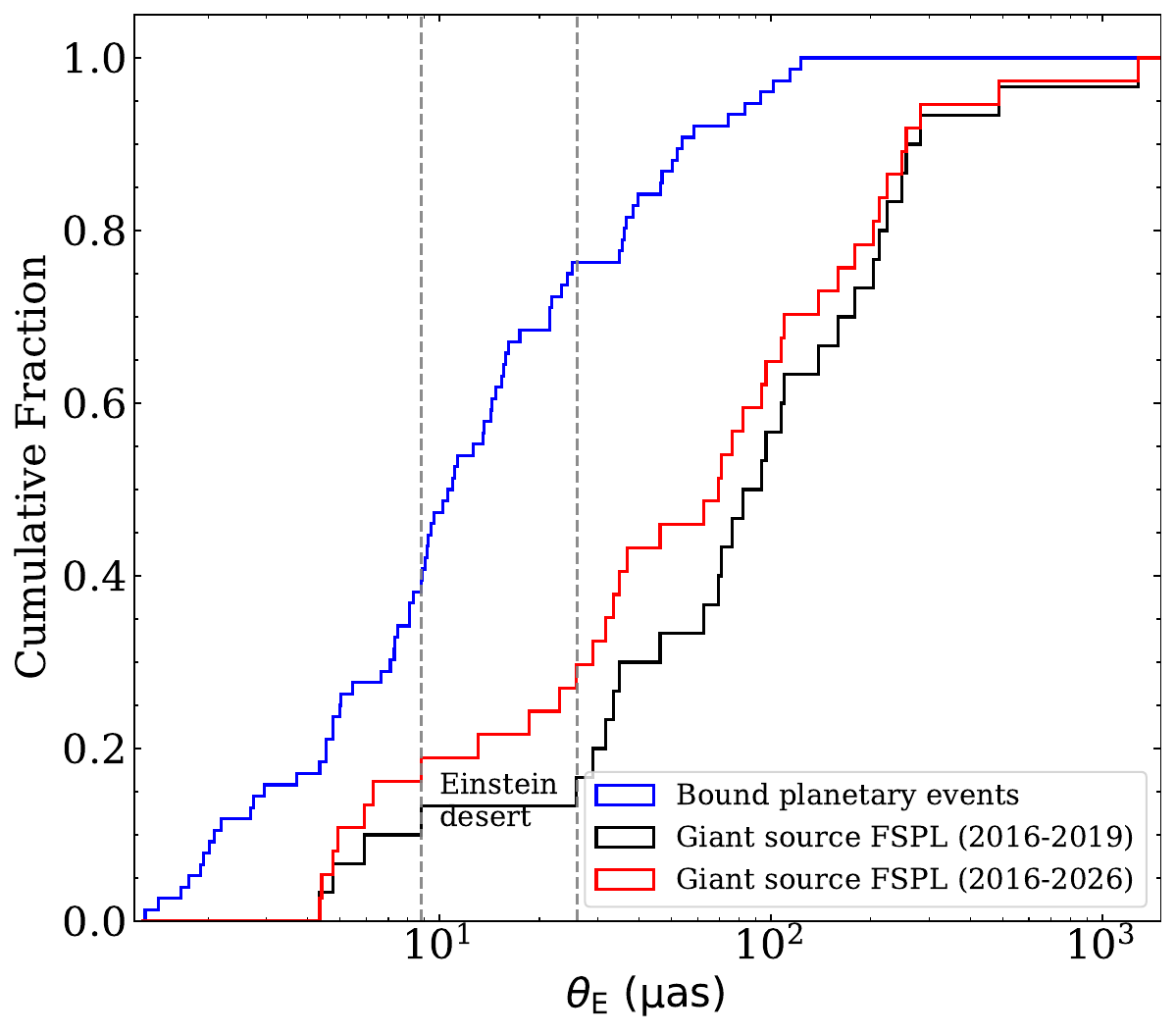}
    \caption{The cumulative distribution of $\thetae$ of isolated objects and bound planetary events detected in KMTNet survey. 
    The grey dash lines show the Einstein desert defined in \cite{Gould2022_FFP_EinsteinDesert}, within which their sample (black curve) is flat. After publication, several events within the desert are detected. The red curve is the combination of \cite{Gould2022_FFP_EinsteinDesert} sample and 5 published giant-source FSPL events from 2020-2026 KMTNet survey. The blue curve is the planetary $\thetae$ distribution from a bound planet sample, which includes the complete sample of 2016--2019 \citep{2017_subprime}, 2021 prime-field events \citep{2021_prime}, and 2023 AlertFinder events \citep{KMT2023_mass1,KMT2023_mass2,KMT2023_mass3}.
    } 
    \label{fig:thetaelist}
\end{figure}

Figure~\ref{fig:thetaelist} compares the evolving KMTNet FSPL sample. The black curve shows the cumulative distribution of the complete sample of 30 giant-source FSPL events identified in the 2016--2019 KMTNet data presented by \citet{Gould2022_FFP_EinsteinDesert}, which exhibited an empty interval at $9~\muas<\thetae<26~\muas$. The red curve includes the \citet{Gould2022_FFP_EinsteinDesert} sample, the published giant-source FSPL events (five in total) identified in the 2020--2026 KMTNet data, and the two events presented in this work. It is important to note that the 2016--2026 sample is not complete, because the giant-source FSPL events from the 2020--2026 KMTNet data have been identified through sporadic publications rather than through a complete search of the dataset. The five events added to the \citet{Gould2022_FFP_EinsteinDesert} sample exhibit a strong bias toward small $\thetae$, with the largest $\thetae=36.5~\mu{\rm as}$ obtained from the event \eventb\ presented in this work. Even so, the red curve shows that the Einstein desert is no longer a flat interval: three subsequent detections, including \eventa, produce a slow but nonzero increase through the nominal desert.

To investigate whether this deficit is a generic feature of the substellar mass regime or instead associated with apparently isolated lenses, we compare the FSPL sample with bound planetary events. We select KMTNet planetary events identified by the KMTNet AnomalyFinder search \citep{OB191053,2019_prime}, including the complete sample of 2016--2019 planets summarized by \cite{OB160007} and \cite{2017_subprime}, the complete sample of 2021 prime-field events \citep{2021_prime}, and the complete sample of 2023 AlertFinder planets \citep{KMT2023_mass1,KMT2023_mass2,KMT2023_mass3}, for a total of 153 events. For a lens system consisting of a host and a planet with a mass ratio $q=M_{\rm p}/M_{\rm host}$, we define the Einstein radius associated with the planetary mass as
\begin{equation}
\theta_{{\rm E},{\rm p}}=\left(\frac{q}{1+q}\right)^{1/2}\theta_{{\rm E},{\rm tot}},
\end{equation}
where $\theta_{{\rm E},{\rm tot}}$ is the Einstein radius corresponding to the total lens mass. In this sample, we only include events for which $\theta_{{\rm E},{\rm tot}}$ is constrained at the $\geq3\sigma$ level. For events with degenerate solutions, we adopt $q$ and $\theta_{{\rm E},{\rm tot}}$ from the best-fit solution. This selection yields a sample of 78 bound planets, for which the $\theta_{{\rm E},{\rm p}}$ cumulative distribution is shown by the blue curve in Figure~\ref{fig:thetaelist}.

In contrast to the underdensity in isolated-lens distribution, the bound planet cumulative distribution grows rapidly in $9~\muas<\theta_{{\rm E},{\rm p}}<26~\muas$. Note that the three curves are normalized separately, this comparison concerns their shapes rather than their absolute occurrence rates. Moreover, the bound-planet curve is not a complete census because a substantial number of published planetary events lack useful $\thetae$ measurements. However, this preliminary comparison still highlights the different behavior of the distribution of bound and free-floating substellar populations. 

If the contrast between the two distributions remains after accounting for their selection functions, it would suggest that the Einstein desert is not simply inherited from the formation distribution of planets in protoplanetary disks. Instead, it may be related to the mass dependence of the processes that create an isolated population. For example, dynamical interactions may preferentially populate the isolated low-mass population through scattering and ejection from planetary systems \citep{Rice2003,Stamatellos2009,Ma2016}, while objects above the desert may receive an increasing contribution from star-like fragmentation and gravitational collapse \citep{Padoan2004,Bate2012,Luhman2012ARAA}. In this picture, the desert would mark a transition between populations produced predominantly through different channels. The interpretation is not unique because the ejection efficiency depends on the system architecture and dynamical history, and the two formation channels may overlap substantially in mass.

Historically, events below the lower shore of the Einstein desert, with $\thetae\lesssim9~\muas$, have been securely interpreted as planetary-mass isolated objects, although the boundary of the FFP population has not been well established. Even if the upper shore of the Einstein desert is adopted as the boundary, our bound-planet sample shows that $\sim23\%$ of the events lie above it. Equations~\ref{eq:thetaE_M_a} and \ref{eq:thetaE_M_b} likewise show that the angular Einstein radii of the two FSPL events presented in this paper, which lie slightly above the upper shore, are compatible with giant-planet masses for typical lens and source distances. A more complete understanding of the FFP population therefore requires direct lens-mass measurements. These can be obtained through simultaneous space- and ground-based observations that enable satellite microlensing parallax measurements. Combining the parallax $\pie$ with the angular Einstein radius $\thetae$ yields the lens mass, $M=\thetae/(\kappa\pie)$ \citep{Gould2000}. For example, by combining ground-based KMTNet and OGLE observations with Gaia observations, \cite{Dong2026Sci} directly measured the lens mass of KMT-2024-BLG-0792, finding a Saturn-mass lens with $\thetae=18.6 \pm 0.9~ \muas$. The forthcoming Galactic Bulge Time-Domain Survey (GBTDS) of the Nancy Grace Roman Space Telescope \citep{MatthewWFIRSTI} and the Earth 2.0 (ET) Microlensing Telescope \citep{CMST,ET}, specifically designed for satellite parallax measurements, are expected to directly measure the masses of hundreds of FFPs, thereby providing a more complete understanding of the population.

\bigskip

Q.Q., W.Zang, H.Y., S.M., J.Z., Z.L., Y.T. and W.Zhu acknowledge support by the National Natural Science Foundation of China (Grant No. 12133005, PI: S.M.). This research has made use of the KMTNet system operated by the Korea Astronomy and Space Science Institute (KASI) at three host sites of CTIO in Chile, SAAO in South Africa, and SSO in Australia. Data transfer from the host site to KASI was supported by the Korea Research Environment Open NETwork (KREONET). This research was supported by KASI under the R\&D program (project No. 2026-1-904-01) supervised by the Ministry of Science and ICT. The MOA project is supported by JSPS KAKENHI Grant Number 
JP16H06287, JP22H00153, JP23KK0060 and JP25H00668. The OGLE project has received funding from the Polish National Science Centre grant OPUS 2024/55/B/ST9/00447 awarded to A.U. This work makes use of observations from the Las Cumbres Observatory global telescope network. This work is part of the ET space mission which is funded by the China's Space Origins Exploration Program. J.C.Y. acknowledges support from U.S. NASA Grant No. 80NSSC25K7146. J.C.Y. acknowledges support from a Scholarly Studies grant from the Smithsonian Institution. Work by C.H. was supported by the grants of National Research Foundation of Korea (2019R1A2C2085965 and 2020R1A4A2002885). This research was funded in part by National Science Centre, Poland, grant SONATA 2023/51/D/ST9/00187 awarded to P.M. RP acknowledges support by the National Science Center, Poland, grant Sonata Bis 2021/42/E/ST9/00038. The authors acknowledge the High-performance Computing Center at Westlake University for providing computational and data storage resources that have contributed to the research results reported within this paper. 

This project used data obtained with the Dark Energy Camera (DECam), which was constructed by the Dark Energy Survey (DES) collaboration. Funding for the DES Projects has been provided by the U.S. Department of Energy, the U.S. National Science Foundation, the Ministry of Science and Education of Spain, the Science and Technology Facilities Council of the United Kingdom, the Higher Education Funding Council for England, the National Center for Supercomputing Applications at the University of Illinois at Urbana-Champaign, the Kavli Institute for Cosmological Physics at the University of Chicago, the Center for Cosmology and Astro-Particle Physics at The Ohio State University, the Mitchell Institute for Fundamental Physics and Astronomy at Texas A\&M University, Financiadora de Estudos e Projetos, Funda\c{c}\~ao Carlos Chagas Filho de Amparo \`a Pesquisa do Estado do Rio de Janeiro, Conselho Nacional de Desenvolvimento Cient\'{\i}fico e Tecnol\'ogico and the Minist\'erio da Ci\^encia, Tecnologia e Inova\c{c}\~ao, the Deutsche Forschungsgemeinschaft, and the collaborating institutions in the Dark Energy Survey.

The collaborating institutions are Argonne National Laboratory; the University of California at Santa Cruz; the University of Cambridge; Centro de Investigaciones Energ\'eticas, Medioambientalesy Tecnol\'ogicas (CIEMAT), Madrid; the University of Chicago; University College London; the DES-Brazil Consortium; the University of Edinburgh; the Eidgen\"ossische Technische Hochschule (ETH) Z\"urich; Fermi National Accelerator Laboratory; the University of Illinois at Urbana-Champaign; the Institut de Ci\`encies de l'Espai (IEEC/CSIC); the Institut de F\'{\i}sica d'Altes Energies (IFAE); Lawrence Berkeley National Laboratory; the Ludwig-Maximilians-Universit\"at M\"unchen and the associated Excellence Cluster Universe; the University of Michigan; NSF NOIRLab; the University of Nottingham; The Ohio State University; the OzDES Membership Consortium; the University of Pennsylvania; the University of Portsmouth; SLAC National Accelerator Laboratory; Stanford University; the University of Sussex; and Texas A\&M University.

This project used data from the DECam Rogue Earths and Mars Survey (DREAMS), whose primary light-curve generation and archive are hosted by the Department of Astronomy at Westlake University. Based on observations at NSF Cerro Tololo Inter-American Observatory, NSF NOIRLab (NOIRLab Prop.\ ID 2025A-806294, PI: Weicheng Zang; Prop.\ ID 2025B-560332, PI: Weicheng Zang \& Hongjing Yang), which is managed by the Association of Universities for Research in Astronomy (AURA) under a cooperative agreement with the U.S. National Science Foundation.

We used ChatGPT (OpenAI) to assist with language editing. The authors are responsible for the scientific results and the final manuscript.

\software{pySIS \citep{pysis,Yang_TLC,Yang_TLC2}, numpy \citep{numpy}, emcee \citep{emcee2,emcee}, Matplotlib \citep{Matplotlib}, SciPy \citep{scipy}}, VBBL \cite{Bozza2010,Bozza2018,VBMicrolensing2025}

\bibliography{ref.bib}

\end{CJK*}
\end{document}